# Fuzzy Decision Analysis in Negotiation between the System of Systems Agent and the System Agent in an Agent-Based Model

Paulette Acheson, Cihan Dagli

Engineering Management & Systems Engineering
Department
Missouri University of Science & Technology
MO, USA
pbatk5@mail.mst.edu, dagli@mst.edu

Nil Kilicay-Ergin

Great Valley School of Graduate Professional Students
Penn State University
PA, USA
nhe2@psu.edu

*Abstract*— **Previous papers have described a computational approach to System of Systems (SoS) development using an Agent-Based Model (ABM). This paper describes the Fuzzy Decision Analysis used in the negotiation between the SoS agent and a System agent in the ABM of an Acknowledged SoS development. An Acknowledged SoS has by definition a limited influence on the development of the individual Systems. The individual Systems have their own priorities, pressures, and agenda which may or may not align with the goals of the SoS. The SoS has some funding and deadlines which can be used to negotiate with the individual System in order to illicit the required capability from that System.**

**The Fuzzy Decision Analysis determines how the SoS agent will adjust the funding and deadlines for each of the Systems in order to achieve the desired SoS architecture quality. The Fuzzy Decision Analysis has inputs of performance, funding, and deadlines as well as weights for each capability. The performance, funding, and deadlines are crisp values which are fuzzified. The fuzzified values are then used with a Fuzzy Inference Engine to get the fuzzy outputs of funding adjustment and deadline adjustment which must then be defuzzified before being passed to the System agent.**

**The first contribution of this paper is the fuzzy decision analysis that represents the negotiation between the SoS agent and the System agent. A second contribution of this paper is the method of implementing the fuzzy decision analysis which provides a generalized fuzzy decision analysis.**

## I. INTRODUCTION

The computational approach to System of Systems (SoS) development using an Agent-Based Model (ABM) was presented in previous papers [1][2]. These papers provided the mathematical representation of SoS development based on the Wave Model [3]. The research included the development of a prototype model of an Acknowledged SoS development [1]. One thing that was absent from the previous works was the negotiation between the SoS agent and the System agent. This paper contributes the fuzzy decision analysis within the SoS agent that supports the negotiation between the SoS agent and the System agent.

## II. BACKGROUND

According to the DoD, an Acknowledged SoS has a specific definition [4]. An Acknowledged SoS is the type of SoS where some organization, such as an SoS manager, is responsible for the SoS development and can exert some minimal influence on the development of the individual systems that comprise the SoS. The SoS manager has a set of SoS capabilities and deadlines for those capabilities and some funding for development of those SoS capabilities. The SoS manager can use the funding and deadlines to influence the individual systems to provide capabilities to the SoS.

The previous research [1][2] describes the Agent-Based Model (ABM) that was implemented in AnyLogic [5]. The agent-based paradigm is based on object-oriented concepts [6] derived from software engineering. The ABM was developed using an Object-Oriented System Architecture approach as described in [7]. The ABM has a user interface where an SoS acquisition officer could run different scenarios. The ABM has an SoS agent and a System agent where the SoS agent is instantiated once and the System agent is instantiated ten times, representing the SoS that is comprised of ten systems. Future research will allow the user to set the number of systems in the SoS. In the ABM simulation the one instance of the SoS is simultaneously negotiating with all ten instances of the individual systems.

As mentioned in [2], the SoS development starts with an initial SoS architecture that supports a set of capabilities (Ci). Each capability (Ci) has an associated weight (wi). The SoS agent has a set of desired capabilities and associated performance levels and deadlines for each system. In order to encourage the system to provide the desired capabilities and performance levels at the requested deadlines, the SoS agent also offers some additional funding to the system. The system has its own priorities, schedules, funding, and goals and returns to the SoS agent what performance levels and deadlines it can support for what amount of funding. This paper focuses on the negotiation model within the SoS agent. So for the purpose of this paper the system is a black box that receives capabilities, performance levels, deadlines, and funding and then returns an updated set of performance levels, deadlines, and funding. The SoS agent must then determine if the results from all the systems are acceptable or if further negotiation is required. In order to determine if the results from all the systems are acceptable, the SoS agent uses a different Fuzzy Assessor to qualitatively evaluate the overall SoS architecture. This Fuzzy Assessor is described in [1] and [2] and is not part of this paper.

The SoS agent is influenced by environmental factors of National Priorities and Threats. In defense spending, the National Priorities can influence which programs get the funding and support to finish development. Similarly, the DoD is most likely to support and fund those programs that address the current Threats. The ABM allows the user to set the National Priorities and Threats. This effectively sets the weight for each capability similar to what occurs in real world SoS development where National Priorities dictate which capabilities get the most attention and funding.

### III. APPROACH

This paper addresses the negotiation model that takes place within the SoS agent when the actual performance levels, deadlines, and capabilities from the individual systems do not provide an acceptable overall SoS architecture quality level. If the systems provided exactly what the SoS agent requested, then there would be no need for negotiation. But in most SoS development, the resulting capabilities from the systems are rarely what the SoS requested.

The difference between the performance requested by the SoS and the performance returned by the system is called the Performance Gap. Performance Gap has the fuzzy values of {None, Low, High, Extreme}. The difference between the deadline requested by the SoS and the deadline returned by the system is called the Deadline Gap. Deadline Gap has fuzzy values of {None, Low, High, Extreme}. The difference between the funding provided by the SoS and the funding requested by the system is called the Funding Gap. Funding Gap has fuzzy values of {None, Low, High, Extreme}. The Weight for each capability has fuzzy values of {None, Low, High, Heavy}. Figure 1 shows the fuzzy membership functions for Performance Gap.

The negotiation assumes that capabilities with higher weights are more important and deserve more funding than capabilities with lower weights. Thus when there is a gap between requested and actual performance, the negotiation will move funding from lower weighted capabilities to higher weighted capabilities.

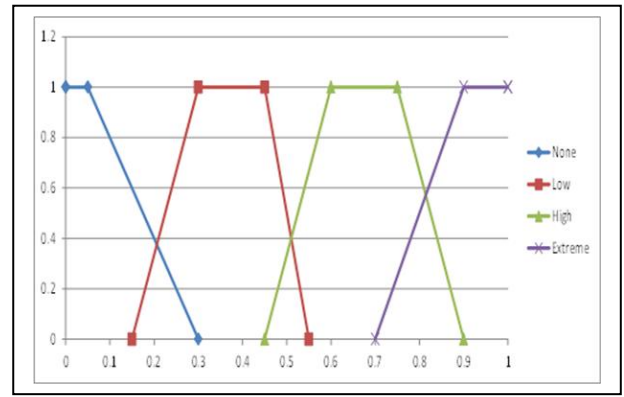

Figure 1.  Performance Gap Fuzzy Membership Function.

When the capabilities are equally weighted, the negotiation assumes that it is better to have all the capabilities at lower performance levels and at longer deadlines than to be missing some capabilities. Thus when the capabilities are equally weighted, the negotiation will move funding and deadlines toward having all capabilities but less than perfect performance levels. This means that funding will be moved from capabilities that have a Performance Gap=None and toward capabilities that have a Performance Gap=High or Extreme. Similarly, the deadlines will be increased for capabilities that have a Performance Gap=High or Extreme and decreased for capabilities that have a Performance Gap=None. This approach pushes Performance Gap toward the middle values.

### IV. FUZZY DECISION ANALYSIS

The first contribution of this paper is the fuzzy decision analysis which represents the SoS agent side of the negotiation between the SoS agent and the System agent. This analysis resides inside the SoS agent and is expressed as an Fuzzy Negotiation. The Fuzzy Negotiation model is a fuzzy decision analysis and is illustrated in Figure 2.

The Performance Gap, Funding Gap, and Deadline Gap are differences between what the SoS is requesting and what the system agrees to provide. These are crisp values that must first be fuzzified before used as inputs to the Fuzzy Negotiation. The deadlines represent the Wave cycle that will implement the capability. These are integers and range from 1-100. The funding represents millions of dollars and these are real numbers that range from 1-10. The performance value is an integer that ranges from 1-10 where 10 means all the performance is present and 1 means there is no performance. The fuzzification of the Performance Gap, Funding Gap, and Deadline Gap is presented in Table 1.

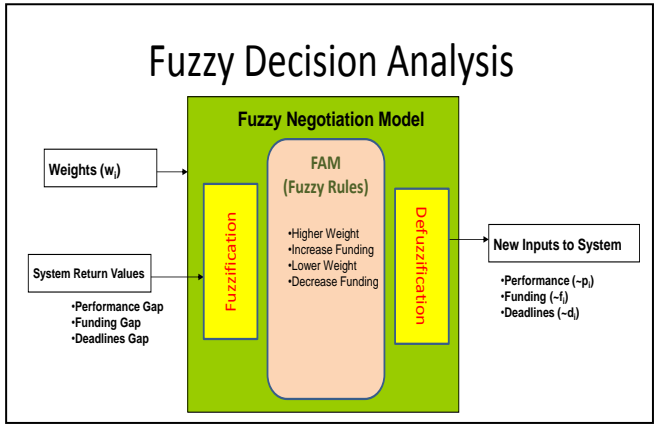

Figure 2.  Fuzzy Decision Analysis with Fuzzy Negotiation Model

TABLE I.        FUZZIFICATION OF CRISP INPUTS TO FUZZY NEGOTIATION

|  | None | Low | High | Extreme |
|---|---|---|---|---|
| Performance Gap | 0 | < 2 | < 7 and > 2 | > 7 |
| Funding Gap | 0 | < 3.5 | < 6.5 and > 3.5 | > 6.5 |
| Deadline Gap | 0 | < 20 | < 80 and > 20 | > 80 |

The Fuzzy Negotiation model is a set of Fuzzy Membership functions and a Fuzzy Inference Engine implemented in a Fuzzy Associative Memory (FAM). The inputs to the Fuzzy Inference Engine are the Performance Gap, Funding Gap, Deadline Gap, and the Weight for each capability. The outputs of the Fuzzy Inference Engine are the Funding Adjustment and Deadline Adjustment. The rules for the Fuzzy Inference Engine are implemented as a FAM. Since there are four inputs there are 256 rules in this FAM so not all the rules are listed in this paper. In order to help explain the Fuzzy Negotiation, a portion of these rules are shown in Table 2.

The FAM provides two fuzzy outputs, namely Funding Adjustment and Deadline Adjustment. A centroid defuzzification is used on these fuzzy outputs to produce the crisp values that are then sent to the system.

TABLE II.        PARTIAL SET OF FUZZY RULES IN FAM

| Inputs | | | | Outputs | |
|---|---|---|---|---|---|
| Performance Gap | Weight | Funding Gap | Deadline Gap | Funding | Deadline |
| none | none | none | none | decrease little | do nothing |
| Low | none | none | none | do nothing | shorten |
| High | none | none | none | increase little | extend |
| extreme | none | none | none | increase little | big delay |
| none | low | none | none | decrease little | do nothing |
| Low | low | none | none | do nothing | shorten |
| High | low | none | none | increase little | extend |
| extreme | low | none | none | increase little | big delay |
| none | high | none | none | do nothing | do nothing |
| Low | high | none | none | do nothing | do nothing |
| High | high | none | none | increase much | do nothing |
| extreme | high | none | none | increase much | do nothing |
| none | heavy | none | none | do nothing | do nothing |
| Low | heavy | none | none | increase little | do nothing |
| High | heavy | none | none | increase much | do nothing |
| extreme | heavy | none | none | increase much | do nothing |

## V.    IMPLEMENTATION

The second contribution of this paper is the method of fuzzy decision implementation. The Fuzzy Decision uses the Fuzzy Associative Memory (FAM) which is read in from a file during initialization. The Fuzzy Inference Rules can be specified by the user via this input file. This implementation provides a generalized fuzzy decision implementation that can be adapted to any domain.

The implementation stores the FAM as a multidimensional array in memory. The fuzzy values for Performance Gap, Weight, Funding Gap, and Deadline Gap are used as enumerated values to index into the FAM. For example, the following fuzzy rule in Figure 3 would be indexed into the FAM as shown in Figure 4.

> If Performance Gap is high and Weight is heavy and Funding Gap is none, then adjustment to Funding is much increase.

Figure 3.   Typical Fuzzy Rule

FAM(high, heavy, none) = much increase

Figure 4.   Example Indexing into FAM

During initialization, the set of all fuzzy rules are read from an Excel spreadsheet and stored into the FAM. By implementing a FAM and reading the rules from a file, the model is a generalized structure for simulating any Fuzzy Decision Analysis. The actual results and execution as specified in [1] and [2] were based on an ISR domain but the model works equally well for another SoS domain, such as a Space domain by merely changing the fuzzy rules in the input file to reflect the Space domain. The use of a FAM also allows future versions of the model the option of adjusting the rules during execution.

## VI.   INTEGRATION

This research was part of a project to model and simulate the SoS development which included other models developed by different people. These other models were developed in Matlab [8] while the agent-based model which includes the fuzzy decision analysis was developed in AnyLogic [5]. Since Matlab was based on the C-language and AnyLogic was based on the Java language, the challenge was how to integrate models developed in these different tools.

The best solution was determined to be the use of Matlab executables called from within the agent-based model. Matlab code provided in *.m files was compiled and linked within Matlab using the "mcc –m" command. The interface between the different executables was accomplished through the use of Excel files for inputs and outputs. As long as the Matlab run time library was installed on the computer running the simulation, the Matlab executables ran fine. This allowed users to run the simulation without requiring a Matlab license.

The use of the Excel file interface made the simulation run slower but provided clear delineation between the executables. This approach to integration allowed each person on the project team to develop their model independently of the others and still be able to run the overall set of models as one cohesive simulation.

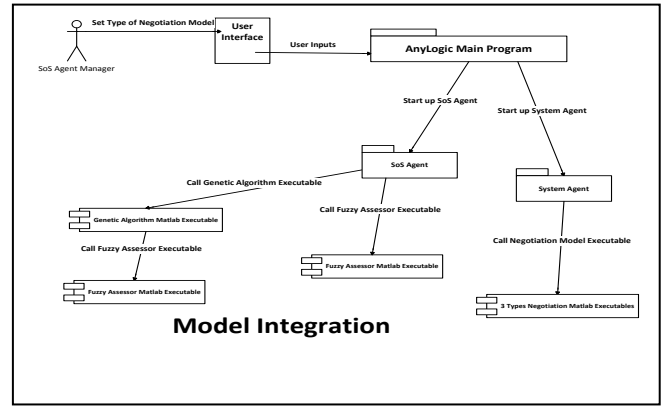

Figure 5.   Model Integration Structure

## VII.   CONCLUSION AND FUTURE WORK

This paper presents a fuzzy decision analysis that represents the fuzzy negotiation model within the SoS agent. The fuzzy negotiation model is part of the ABM representing Acknowledged SoS development. The use of a FAM to implement the Fuzzy Inference Engine rules provides advantages over implementation methods and is the second contribution of this paper.

Future work would involve adding the capability for the fuzzy decision analysis to adjust the fuzzy rules as the model runs and "learns" as it executes. Future work also includes methods of adjusting the fuzzy membership functions during execution. This would add an Intelligence Model to the SoS agent.

The agent-based model was developed in AnyLogic which is Java based. It is possible to create Java applets from Java code. Future work would be to create a Java applet that would run the set of models and not require an AnyLogic license.

### ACKNOWLEDGMENT

This material is based upon work supported, in whole or in part, by the U.S. Department of Defense through the Systems Engineering Research Center (SERC) under Contract H98230-08-D-0171. SERC is a federally funded University Affiliated Research Center managed by Stevens Institute of Technology.

Any opinions, findings and conclusions or recommendations expressed in this material are those of the author(s) and do not necessarily reflect the views of the United States Department of Defense.

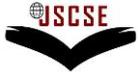